\documentclass[aps,prb,twocolumn,superscriptaddress, longbibliography]{revtex4-2}
\usepackage{graphicx}
\usepackage{latexsym}
\usepackage{amssymb}
\usepackage{amsmath}
\usepackage{amsfonts}
\usepackage[vcentermath]{youngtab}
\usepackage{upgreek}
\usepackage{bm,dsfont}
\usepackage{multirow}
\usepackage{enumerate}
\usepackage[usenames, dvipsnames]{color}
\usepackage[svgnames]{xcolor}
\usepackage{hyperref}
\hypersetup{
colorlinks = true,
linkcolor = [rgb]{0.70,0.13,0.13}, 
citecolor = [rgb]{0.13,0.55,0.13},
urlcolor = [rgb]{0.25, 0.41, 0.88}}
\usepackage{makecell}

\newcommand{\ket}[1]{|#1 \rangle}

\newcommand{\dd}{\mathrm{d}}

\newcommand{\ii}{\mathrm{i}}
\newcommand{\e}{\mathrm{e}}

\newcommand{\eV}{\,\mathrm{eV}}
\newcommand{\U}{\mathrm{U}}
\newcommand{\SU}{\mathrm{SU}}

\newcommand{\Spin}{\mathrm{Spin}}

\newcommand{\dsZ}{\mathbb{Z}}

\newcommand{\scH}{\mathcal{H}}

\newcommand{\Tr}{\operatorname{Tr}}

\newcommand{\vect}[1]{{\bm{#1}}}

\newcommand{\eq}[1]{\begin{equation}#1\end{equation}}
\newcommand{\eqs}[1]{\begin{equation}\begin{split}#1\end{split}\end{equation}}
\newcommand{\eqnref}[1]{Eq.\,\eqref{#1}}
\newcommand{\figref}[1]{Fig.\,\ref{#1}}
\newcommand{\tabref}[1]{Tab.\,\ref{#1}}
\newcommand{\secref}[1]{Sec.\,\ref{#1}}

\newcommand{\refcite}[1]{Ref.\,\onlinecite{#1}}

\newcommand{\topic}[1]{}

\begin{document}

\title{Superconductivity from Doping Symmetric Mass Generation Insulators:\\ Application to La$_3$Ni$_2$O$_7$ under Pressure}

\author{Da-Chuan Lu}
\affiliation{Department of Physics, University of California, San Diego, CA 92093, USA}
\author{Miao Li}
\affiliation{School of Physics, Zhejiang University, Hangzhou 310027, China}
\author{Zhao-Yi Zeng}
\affiliation{School of Physics, Fudan University, Shanghai 200438, China}
\author{Wanda Hou}
\affiliation{Department of Physics, University of California, San Diego, CA 92093, USA}
\author{Juven Wang}
\affiliation{Center of Mathematical Sciences and Applications, Harvard University, Cambridge, Massachusetts 02138, USA}
\author{Fan Yang}
\affiliation{School of Physics, Beijing Institute of Technology, Beijing 100081, China}
\author{Yi-Zhuang You}
\affiliation{Department of Physics, University of California, San Diego, CA 92093, USA}

\begin{abstract}
We investigate the bilayer nickelates as a platform to realize the symmetric mass generation (SMG) insulator, a featureless Mott insulator that arises due to the Lieb-Schultz-Mattis (LSM) anomaly cancellation in bilayer spin-1/2 lattice systems. Through a single-orbital bilayer square lattice model involving intralayer hopping $t$ and interlayer superexchange interaction $J$, we demonstrate the emergence of high-temperature superconductivity (SC) upon doping the SMG insulator. The SC phase features $s$-wave interlayer spin-singlet pairing and exhibits a crossover between the Bardeen-Cooper-Schrieffer (BCS) and 
Bose-Einstein condensation (BEC) limits by tuning the $J/t$ ratio. We estimate the SC transition temperature $T_c$ from both the weak and strong coupling limits at the mean-field level. Our findings offer insights into the experimentally observed decrease in $T_c$ with pressure and the strange metal behavior above $T_c$. Additionally, we propose that both Ni $3d_{z^2}$ and $3d_{x^2-y^2}$ orbitals can exhibit superconductivity in La$_3$Ni$_2$O$_7$ under pressure, but their $T_c$  should vary in opposite ways under doping. This characteristic difference suggests a potential experimental pathway to identify which electronic orbital plays the principal role in the formation of superconductivity in this system.
\end{abstract}

\maketitle

\section{Introduction}
The recent experimental discovery \cite{Sun2023S2305.09586, Liu2023E2307.02950, Hou2023E2307.09865, Zhang2023H2307.14819} of high-temperature superconductivity (SC) in single crystals of La$_3$Ni$_2$O$_7$ under high pressure has uncovered an exciting platform for investigating and testing unconventional SC mechanisms \cite{Lee2006D, Keimer2015F} in condensed matter physics. This nickelate compound, when driven into a pressure-induced orthorhombic phase with $Fmmm$ spacegroup symmetry, manifests superconductivity, with an observed maximum transition temperature ($T_c$) of 80K within a pressure range of 14.0 to 43.5GPa \cite{Sun2023S2305.09586}.

Density-functional-theory (DFT) based first-principles calculations indicate that the low-energy behavior in La$_3$Ni$_2$O$_7$ is governed by two $e_g$ orbitals of Ni: $3d_{x^2-y^2}$ and $3d_{z^2}$ \cite{Sun2023S2305.09586, Luo2023B2305.15564, Zhang2023E2306.03231, Christiansson2023C2306.07931, Shilenko2023C2306.14841, Sakakibara2023P2306.06039}. Within a unit cell, the two $d_{z^2}$ orbitals in distinct layers can couple through hybridization with the apical oxygen's $p$-orbital in the charge-reservoir layer. When subjected to pressure, the Ni-O-Ni bonding angle shifts from 168$^{\circ}$ to 180$^{\circ}$, remarkably enhancing the interlayer coupling. The emergence of SC solely under pressure implies that this interlayer coupling is likely pivotal to the pairing mechanism. Additionally, the robust electron-electron interaction in the Ni-$3d$ orbitals could drive the SC, in line with recent experimental findings that La$_3$Ni$_2$O$_7$ is in proximity to a Mott phase and exhibits non-Fermi-liquid behaviors \cite{Liu2023E2307.02950, Zhang2023H2307.14819}.

Two remarkable features of the La$_3$Ni$_2$O$_7$ superconductor stand out: First, its $T_c$ intriguingly decreases with increasing pressure, a behavior challenging the standard phonon or magnon-mediated pairing mechanism, since higher pressures typically imply stronger lattice vibrations and magnetism energy scales, which would otherwise lead to higher $T_c$. Secondly, above $T_c$, the compound displays a ``strange metal'' behavior, characterized by a linear temperature-dependent resistivity up to 300K \cite{Sun2023S2305.09586, Liu2023E2307.02950, Zhang2023H2307.14819}, which suggests the presence of strong quantum fluctuations or proximity to quantum critical points.

The theoretical framework to comprehensively explain these intriguing features is still under development. On the weak-coupling side, studies employing the functional renormalization group \cite{Yang2023P2306.03706, Gu2023E2306.07275}, fluctuation-exchange approach \cite{Sakakibara2023P2306.06039}, and random-phase approximation \cite{Liu2023T2307.10144, Zhang2023S2307.15276, Zhang2023T2308.07386, Lechermann2023E2306.05121} have been conducted. The majority of these point to the $s^{\pm}$-wave pairing \cite{Yang2023P2306.03706, Gu2023E2306.07275, Sakakibara2023P2306.06039, Liu2023T2307.10144, Zhang2023S2307.15276, Zhang2023T2308.07386}, driven by spin fluctuations. In these models, the nesting between the emergent $\gamma$-pocket under pressure and the previously existing $\beta$-pocket plays a crucial role in the $s^{\pm}$-wave pairing. On the strong-coupling side, various models have been established, involving different types of superexchange interactions \cite{Yang2023P2306.03706, Gu2023E2306.07275, Sakakibara2023P2306.06039, Shen2023E2306.07837,  Wu2023C2307.05662, Cao2023F2307.06806, Lu2023I2307.14965, Oh2023T2307.15706, Qu2023B2307.16873, Yang2023M2308.01176, Liao2023E2307.16697, Jiang2023H2308.06771}. Although most emphasize the significance of the interlayer exchange between the two $d_{z^2}$ orbitals within a unit cell, differing opinions exist on the dominant orbital contributing to SC, which can be either $d_{z^2}$ \cite{Sakakibara2023P2306.06039, Shen2023E2306.07837, Yang2023M2308.01176} or $d_{x^2-y^2}$ \cite{Lu2023I2307.14965, Oh2023T2307.15706, Qu2023B2307.16873, Jiang2023H2308.06771}. Currently, a consensus has not been reached regarding which orbital contributes more significantly to SC, and a unified physical understanding remains elusive.

Our study aims to establish a simple theoretical model that encapsulates the main features of the pressurized nickelate high-temperature superconductor. Specifically, we propose to understand the high-$T_c$ superconductivity in La$_3$Ni$_2$O$_7$ from the perspective of doped Symmetric Mass Generation (SMG) \cite{Wang2022S2204.14271, Slagle2015E1409.7401, He2016Q1603.08376, You2018S1705.09313, You2018F1711.00863, Lu2023F2210.16304, Hou2022V2212.13364, Lu2023G2307.12223} insulators. One critical distinction between nickelate and cuprate superconductors lies in their lattice structures: nickelates have a bilayer lattice structure with a strong interlayer superexchange, while cuprates typically consist of decoupled single layers. This difference results in fundamentally contrasting Mott insulating states when these systems are at half-filling, particularly in regard to the Lieb-Schultz-Mattis (LSM) \cite{Lieb1961T, Oshikawa2000Tcond-mat/0002392, Cheng2016T1511.02263, Cho2017A1705.03892, Bultinck2018F1808.00324} anomaly.

In the cuprates, the Mott insulator is constrained by the LSM theorem to possess a non-trivial ground state, meaning that the cuprate Mott state must either spontaneously break the symmetry or develop a topological order \cite{Hastings2005Scond-mat/0411094, Arovas2022T2103.12097}. Conversely, nickelates are not restricted by the LSM theorem as the LSM anomaly (at half filling) falls under the $\dsZ_2$ classification. Thus, the bilayer system is anomaly-free and can form a trivial Mott insulator under interaction, which is referred to as an SMG insulator. Doping an SMG insulator also leads to superconductivity, but the underlying mechanisms differ significantly from those of the traditional doped Mott insulator.

\begin{figure}[htbp]
\begin{center}
\includegraphics[scale=0.65]{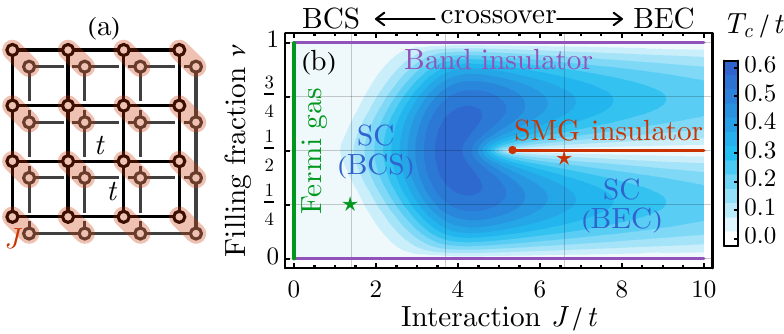}
\caption{(a) The proposed bilayer square lattice model, featuring intralayer hopping $t$ and interlayer Heisenberg interaction $J$. (b) Mean-field phase diagram, dominated by a single SC phase (blue) with $s$-wave interlayer spin-singlet pairing, alongside Fermi gas (green), band insulator (purple), and SMG insulator (red) lines. The shading in blue illustrates the SC transition temperature $T_c$, which crosses over between BCS and BEC regimes depending on $J/t$. Red and green stars highlight parameters relevant to the Ni-$3d_{z^2}$ and $3d_{x^2-y^2}$ electrons in La$_3$Ni$_2$O$_7$, respectively. }
\label{fig: phase}
\end{center}
\end{figure}

More specifically, we propose a bilayer square lattice model \figref{fig: phase}(a) with interlayer antiferromagnetic Heisenberg interaction to demonstrate the key physics of an SMG insulator. A mean-field phase diagram is shown in \figref{fig: phase}(b). The emergent SC in the doped SMG insulator can be interpreted within the theoretical framework of the BCS-BEC crossover \cite{Bloch2008M0704.3011, Zwerger2011T}. The observed decrease in $T_c$ with increasing pressure indicates that the system resides on the Bose-Einstein condensate (BEC) side. Pictorially, the SMG insulator can be viewed as a Mott insulating state \cite{Gersch1963Q, Ma1986S, Fisher1989B} of interlayer spin-singlet Cooper pairs \cite{Bardeen1957M, Bardeen1957T}. The pairing energy for these Cooper pairs stems from the interlayer spin interaction. Upon doping, these preformed Cooper pairs become mobile on the lattice, developing a BEC state at low temperatures, which can be interpreted as an $s$-wave interlayer spin-singlet SC state from the electron's perspective. The BEC-side SC phase exhibits three characteristics:
\begin{enumerate}[(i)]
\item The critical temperature $T_c$ decreases with increasing interlayer Heisenberg interaction, which can typically be realized by increasing the pressure exerted on the material.
\item Above $T_c$, the system enters a pseudo-gap (PG) phase, where phase-incoherent preformed Cooper pairs fluctuate in the background. These low-energy fluctuations scatter the itinerant electrons, leading to a strange metal behavior in the system.
\item The optimal doping (for maximal $T_c$) occurs near 50\% (i.e.,\,1/4 filling of electrons) because the interlayer Cooper pairs have the highest mobility at this level, and can most easily establish the phase coherence that is needed for SC.
\end{enumerate}

Combining with the theoretical analysis of the bilayer square lattice model, evidence suggests that both $d_{z^2}$ and $d_{x^2-y^2}$ orbitals can contribute to superconductivity in La$_3$Ni$_2$O$_7$. However, distinguishing between these two possibilities requires a closer examination of how the critical temperature $T_c$ depends on doping. If superconductivity is primarily governed by the BCS mechanism acting on $d_{x^2-y^2}$ electrons, then $T_c$ will rise with \emph{electron} doping. Conversely, if the BEC mechanism governing $d_{z^2}$ electrons predominates, $T_c$ will increase with \emph{hole} doping. These contrasting behaviors in the doping dependence of $T_c$ offer potential experimental pathways for determining the electronic orbital origin of superconductivity in this system.

The remainder of this paper is structured as follows. We begin with a brief review of the local electronic structure of La$_3$Ni$_2$O$_7$ in \secref{sec: local}, upon which we construct a minimal bilayer square lattice model, as detailed in \secref{sec: model}. The symmetry and anomaly analysis, along with an introduction to the SMG insulator, are discussed in \secref{sec: SMG}. The emergence of SC from doping the SMG insulator is then examined from both weak-coupling (\secref{sec: weak}) and strong-coupling (\secref{sec: strong}) perspectives, with a unification of these results in \secref{sec: crossover}. We reflect on the implications of our findings for nickelate superconductors in \secref{sec: application} and briefly explore other strong correlation effects as a supplement to our model in \secref{sec: Hubbard}. The paper concludes with a summary in \secref{sec: summary}.

\section{Modeling}

\subsection{Local Electronic Structure}
\label{sec: local}

\begin{figure}[htbp]
\begin{center}
\includegraphics[scale=0.65]{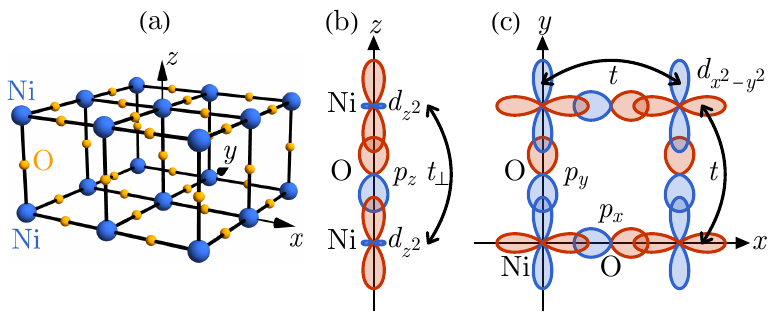}
\caption{(a) Lattice structure of the NiO$_2$ bilayer. (b) Interlayer hopping $t_\perp$ of $d_{z^2}$ electron through the Ni-O-Ni bond. (c) Intralayer hopping $t$ of $d_{x^2-y^2}$ electron trough the Ni-O-Ni bond.}
\label{fig: orbitals}
\end{center}
\end{figure}

The bulk single crystal La$_3$Ni$_2$O$_7$ is constructed from an infinite stack of NiO$_2$ bilayers, see \figref{fig: orbitals}(a). The nickel elements in the compound have a valence of Ni$^{2.5+}$, corresponding to a $3d^{7.5}$ electronic state. Within this state, 6 electrons fully occupy the $t_{2g}$ orbitals ($d_{xz}$, $d_{yz}$, $d_{xy}$), while the remaining 1.5 electrons populate the $e_g$ orbitals ($d_{z^2}$, $d_{x^2-y^2}$). On average, about one electron is on the $d_{z^2}$ orbital (half filling the $d_{z^2}$ band, filling fraction $\nu=1/2$), and on-average 0.5 electron is on the $d_{x^2-y^2}$ orbital (quarter filling the $d_{x^2-y^2}$ band, filling fraction $\nu=1/4$). However, these filling fractions are not fixed, as charge transfer between the two $e_g$ orbitals might occur under pressure.  Density functional theory (DFT) calculations \cite{Luo2023B2305.15564, Zhang2023E2306.03231} indicate that the electronic structure near the Fermi surface is primarily derived from the $d_{z^2}$ and $d_{x^2-y^2}$ orbitals of nickel and the $p_{x/y/z}$ orbitals of oxygen. The dominant hopping is between the $d_{z^2}$ orbitals across the vertical Ni-O-Ni bond, denoted as $t_\perp\approx 0.64\eV$ in \figref{fig: orbitals}(b), and among the $d_{x^2-y^2}$ orbitals across in-plane Ni-O-Ni bonds, denoted as $t\approx 0.48\eV$ in \figref{fig: orbitals}(c). The other hopping processes are found to be parametrically small \cite{Luo2023B2305.15564, Zhang2023E2306.03231}.

Whether the SC arises from the $d_{z^2}$ electron or the $d_{x^2-y^2}$ electron remains a subject of debate. We attempt to identify the unique features of these two possibilities through a unified toy model and compare them with experimental observations. 

Firstly, considering the electron interaction effect, the most prominent local physics within each unit cell is the interlayer antiferromagnetic Heisenberg interaction between $d_{z^2}$ electrons. This arises from the superexchange mechanism across the vertical Ni-O-Ni bond, and the coupling strength is estimated to be of the order $J=4t_\perp^2/U \simeq 0.66 \eV$, assuming a Hubbard interaction around $U\simeq 2.5\eV$. Given that the half-filled $d_{z^2}$ orbital corresponds to two electrons per unit cell, the electrons naturally pair up into interlayer spin-singlets in each unit cell, forming a featureless Mott insulator \cite{Kimchi2013F1207.0498} without symmetry-breaking or topological orders. This insulating state is also referred to as the SMG insulator, defined as a featureless Mott insulator without LSM anomaly (or Fermi surface anomaly), which is only achievable when each unit cell hosts linear representations of internal symmetries (or integer filling of electrons), as is the case here.

Next, given the ferromagnetic Hund's rule interaction between the $d_{x^2-y^2}$ and $d_{z^2}$ electrons within each Ni atom, an interlayer antiferromagnetic Heisenberg interaction between the $d_{x^2-y^2}$ electrons can be induced \cite{Lu2023I2307.14965, Oh2023T2307.15706}, on the $d_{z^2}$ orbital-selective interlayer spin-singlet background \cite{Zhang2023E2306.03231}. The coupling strength $J$ is roughly the same as that for the $d_{z^2}$ electrons \cite{Lu2023I2307.14965, Oh2023T2307.15706}. However, since the $d_{x^2-y^2}$ electron has a quarter filling only, and its intralayer hopping $t\simeq 0.48\eV$ is on a scale close to $J\simeq 0.66\eV$, it has more freedom to hop around the lattice and is expected to be closer to the Fermi liquid state, away from the SMG insulating limit.

Overall, the \(d_{z^2}\) electrons are more localized in the unit cell, exhibiting strong coupling physics; while the \(d_{x^2-y^2}\) electrons are more itinerant on the square lattice, demonstrating weak coupling physics. We conclude that, for both \(d_{z^2}\) and \(d_{x^2-y^2}\) electrons, the competition between interlayer spin-spin interaction $J$ and intralayer hopping $t$ is key to understanding their low-energy physics.

\subsection{Minimal Lattice Model}
\label{sec: model}

Based on this insight, we postulate to ignore other interactions and band structure details, and first focus on the following minimal model on the bilayer square lattice, as depicted in \figref{fig: phase}(a),
\eq{\label{eq: H}
H=t\sum_{\langle ij\rangle,l}(c_{il}^\dagger c_{jl}+\text{h.c.})-\mu\sum_{il}c_{il}^\dagger c_{il}+J\sum_{i}\vect{S}_{i1}\cdot\vect{S}_{i2},}
where $c_{il}=(c_{il\uparrow},c_{il\downarrow})^\intercal$ denotes a spin-1/2 electron annihilation operator on the unit cell $i$ and the layer $l=1,2$, and $\vect{S}_{il}:=\frac{1}{2}c_{il}^\dagger\vect{\sigma} c_{il}$ is the electron spin operator. We ignore the on-site Hubbard interaction for now and will comment on its physical effect later. We do not impose any on-site single-occupancy constraint, such that the electron filling can be adjusted from $\nu=0$ (empty) all the way to $\nu=1$ (four electrons per unit cell).  The model may describe either $d_{z^2}$ or $d_{x^2-y^2}$ electrons depending on the parameter choice. Here, $t$ represents the intralayer (in-plane) hopping, $J$ represents the interlayer spin interaction, and the chemical potential $\mu$ tunes the electron filling. Similar models have previously appeared in the literature \cite{Zhai2009A0905.1711, Zhang2020F2001.09159, Zhang2020D2006.01140, Nikolaenko2021S2103.05009, Bohrdt2021E2107.08043, Bohrdt2022S2108.04118, Hirthe2023M2203.10027,  Lu2023G2307.12223}, and recently proposed to describe the La$_3$Ni$_2$O$_7$ superconductor \cite{Shen2023E2306.07837, Qu2023B2307.16873, Oh2023T2307.15706, Lu2023I2307.14965, Wu2023C2307.05662, Yang2023M2308.01176}. Recent research \cite{Lu2023G2307.12223} has revealed that the bilayer square lattice model features a Fermi surface SMG \cite{Lu2023F2210.16304} phase at half-filling. 

The model respects the bilayer square lattice space group symmetry, including the 2D lattice translation symmetry $\dsZ\times\dsZ$. It also has an internal symmetry of 
\eq{\label{eq: G}
G_\text{general}=\SU(2)\times_{\dsZ_2^\mathrm{F}}(\U(1)_1\times\U(1)_2),} 
which includes the spin $\SU(2)$ symmetry (generated by $\vect{S}=\sum_{il}\vect{S}_{il}$), and two independent charge $\U(1)_l$ symmetries in both layers (generated by $N_l=\sum_{i}c_{il}^\dagger c_{il}$ for $l=1,2$). $\dsZ_2^\mathrm{F}$ indicates the fermion parity symmetry shared between $\SU(2)$ and $\U(1)_1\times\U(1)_2$. The $\U(1)_1\times\U(1)_2$ group may also be rewritten as $\U(1)_+\times_{\dsZ_2^\mathrm{F}}\U(1)_-$ with $\U(1)_\pm$ generated by $N_\pm=N_1\pm N_2$, a more convenient notion for later discussions. 

Given these symmetries, for a generic non-integer filling fraction $\nu$ (i.e.\,$4\nu$ electrons per unit cell), the system is subject to the Fermi surface anomaly \cite{Oshikawa2000Tcond-mat/0002392, Paramekanti2004Econd-mat/0406619, Cheng2016T1511.02263, Cho2017A1705.03892, Bultinck2018F1808.00324, Song2019E1909.08637, Wen2021L2101.08772, Else2021N2007.07896, Else2021S2010.10523, Ma2021E2110.09492, Wang2021E2110.10692, Darius-Shi2022G2204.07585, Cheng2022L2211.12543, Lu2023D2302.12731}, and its ground state must either remain gapless (metallic) or become gapped at the price of developing spontaneous symmetry breaking (SSB) order or topological order.
Mathematically, the classification of Fermi surface anomaly \cite{Lu2023D2302.12731} for a given internal symmetry group
$G \supset \dsZ_2^F$ is governed by the cobordism group \cite{Kapustin1406.7329, Freed1604.06527, Guo1711.11587, Wan1812.11967, Guo1812.11959} ${\rm TP}_1(\Spin \times_{\dsZ_2^\mathrm{F}} G)$.
By taking $G=G_\text{general}$ in \eqnref{eq: G}, we arrive at the following classification 
\eq{{\rm TP}_1(\Spin \times_{\dsZ_2^\mathrm{F}} G_\text{general})=\dsZ\times\dsZ,}
signifying the presence of two perturbative local anomalies, each associated with one $\U(1)$ symmetry. The Fermi surface anomaly is generally non-vanishing unless the filling fractions for both $\U(1)$ symmetries are integers. This situation makes it impossible to trivially gap out Fermion excitations on the Fermi surface, thereby precluding the realization of a featureless insulating state.

\section{Analysis}

\subsection{SMG Insulator at Half-Filling}
\label{sec: SMG}

However, an exception occurs when the system is half-filled ($\nu=1/2$), corresponding to zero chemical potential \(\mu=0\) in the lattice model \eqnref{eq: H} and an average of two electrons per unit cell. In this case, the lattice model possesses a larger internal symmetry 
\eq{\label{eq: G1/2}
G_\text{half-filled}=\SU(2)\times_{\dsZ_2^\mathrm{F}}(\SU(2)_1\times\SU(2)_2),}
promoting both $\U(1)_l$ ($l=1,2$) symmetries to $\SU(2)_l$. The $\SU(2)_l$ groups are generated by $\vect{K}_l=\sum_{i}\vect{K}_{il}$ with $\vect{K}_{il}:=\frac{1}{2}(-)^i \tilde{c}_{il}^\dagger \vect{\sigma} \tilde{c}_{il}$, where $\tilde{c}_{il}=(c_{il\uparrow},c_{il\downarrow}^\dagger)^\intercal$ is the Nambu spinor and $(-)^i$ denotes the stagger sign of momentum $(\pi,\pi)$ on the square lattice. 

Plugging the enlarged symmetry group $G_\text{half-filled}$ in \eqnref{eq: G1/2} into the Fermi surface anomaly classification formula, we obtain 
\eq{{\rm TP}_1(\Spin \times_{\dsZ_2^\mathrm{F}} G_\text{half-filled})=0,}
meaning that the Fermi surface anomaly is trivialized to no anomaly at half-filling. Since $G_\text{general} \subset G_\text{half-filled}$, 
this suggests a trivialization of the anomaly in a subgroup $G_\text{general}$
by an injective pushforward to anomaly-free in a larger group $G_\text{half-filled}$ through an injective (one-to-one) homomorphism $\iota$ via $G_\text{general} \overset{\iota}{\to} G_\text{half-filled}$; 
somehow its novelty is \emph{different} from the Symmetry Extension surjective pullback trivialization mechanism \cite{WangWenWitten1705.06728, Wang1801.05416, Prakash1804.11236, Prakash2011.13921}
which uses a pullback to trivialize an anomaly in a quotient group $G_Q$ to 
anomaly-free in an extended total group $\tilde G$,
under a surjective (onto) homomorphism ${s}$: $\tilde G \overset{s}{\to} G_Q$. Physically, the anomaly cancelation at half-filling is because each unit cell with two electrons can realize the trivial (singlet) representation of all $\SU(2)$ symmetries simultaneously. This enables the fermion system to be gapped into a symmetric product state, permitted by the LSM theorem. 

However, the three $\SU(2)$ symmetries are too restrictive, that they rule out all possible Fermi bilinear gapping terms, as enumerated in \tabref{tab: bilinear}. Therefore, symmetric gapping can only be driven by interaction effects intrinsically without any mean-field-level interpretation --- a mechanism known as SMG \cite{Wang2022S2204.14271}. The interlayer Heisenberg interaction in \eqnref{eq: H} is one such symmetric gapping interaction, as to be elaborated later. The resulting gapped state is called an SMG insulator --- a featureless Mott insulator unique to bilayer lattice systems like nickelates, and it can not be realized in single-layer systems such as cuprates due to the Fermi surface anomaly (or the LSM anomaly). 

\begin{table}[htp]
\caption{Classification of all on-site fermion bilinear gapping terms (local order parameters) by symmetry representations, including: intralayer pairings $\Delta_{il}\simeq c_{il}^\intercal \ii\sigma^2 c_{il}$ and density waves $(\Phi_{il},\vect{\Phi}_{il})\simeq (-)^ic_{il}^\dagger \sigma^\mu c_{il}$, interlayer pairings $(\Delta_{i},\vect{\Delta}_{i})\simeq c_{i1}^\intercal \ii\sigma^2\sigma^\mu c_{i2}$ and excitons $(\Phi_i,\vect{\Phi}_i)\simeq (-)^ic_{i1}^\dagger \sigma^\mu c_{i2}$.}
\begin{center}
\begin{tabular}{ccccccc}
\hline
$\SU(2)$ & $\SU(2)_1$ & $\SU(2)_2$ & $\U(1)_1$ & $\U(1)_2$ & Bilinear order & Energy\\
\hline
\multirow{6}{*}{$\mathbf{1}$} & \multirow{2}{*}{$\mathbf{3}$} & \multirow{2}{*}{$\mathbf{1}$} & $\pm 2$ & $0$ & $\Delta_{i1}^\dagger, \Delta_{i1}$ & \multirow{2}{*}{$0$}\\
& & & $0$ & $0$ & $\Phi_{i1}$ & \\
\cline{2-7}
& \multirow{2}{*}{$\mathbf{1}$} & \multirow{2}{*}{$\mathbf{3}$} & $0$ & $\pm2$ & $\Delta_{i2}^\dagger, \Delta_{i2}$ & \multirow{2}{*}{$0$}\\
& & & $0$ & $0$ & $\Phi_{i2}$ & \\
\cline{2-7}
& \multirow{2}{*}{$\mathbf{2}$} & \multirow{2}{*}{$\mathbf{2}$} & $\pm 1$ & $\pm 1$ & $\Delta_{i}^\dagger, \Delta_{i}$ & \multirow{2}{*}{$-3J/4$}\\
& & & $\pm1$ & $\mp 1$ & $\Phi_{i}^\dagger, \Phi_{i}$ & \\
\hline
\multirow{4}{*}{$\mathbf{3}$} & $\mathbf{1}$ & $\mathbf{1}$ & $0$ & $0$ & $\vect{\Phi}_{i1}+\vect{\Phi}_{i2}$ & $+J/2$\\
\cline{2-7}
& $\mathbf{1}$ & $\mathbf{1}$ & $0$ & $0$ & $\vect{\Phi}_{i1}-\vect{\Phi}_{i2}$ & $-J/2$\\
\cline{2-7}
& \multirow{2}{*}{$\mathbf{2}$} & \multirow{2}{*}{$\mathbf{2}$} & $\pm 1$ & $\pm 1$ & $\vect{\Delta}_{i}^\dagger, \vect{\Delta}_{i}$ & \multirow{2}{*}{$+J/4$}\\
& & & $\pm1$ & $\mp 1$ & $\vect{\Phi}_{i}^\dagger, \vect{\Phi}_{i}$ & \\
\hline
\end{tabular}
\end{center}
\label{tab: bilinear}
\end{table}%

The phase diagram of the model Hamiltonian $H$ in \eqnref{eq: H} at half-filling $\nu=1/2$ has been examined by \refcite{Lu2023G2307.12223}, with the key findings being: 
\begin{enumerate}[(i)]

\item In the larger $J/t$ regime, $H$ has a unique symmetric gapped ground state, smoothly deformable from the product state of interlayer spin-singlet state $\prod_{i}(c_{i1\uparrow}^\dagger c_{i2\downarrow}^\dagger-c_{i1\downarrow}^\dagger c_{i2\uparrow}^\dagger)\ket{0}_c$. This state preserves all lattice and internal $\SU(2)\times_{\dsZ_2^\mathrm{F}}(\SU(2)_1\times\SU(2)_2)$ symmetries, and opens a gap (generating a fermion mass) on the Fermi surface, hence called the SMG insulator \cite{Lu2023F2210.16304}. It has no SSB or topological order. Its existence is enabled by anomaly cancellation \cite{Cheng2022L2211.12543, Lu2023D2302.12731} within the bilayer system.

\item In the smaller $J/t$ regime, due to Fermi surface instability at weak coupling, SSB order emerges in the system. \tabref{tab: bilinear} summarizes all possible local SSB orders. The last column lists the mean-field energy of condensing each order parameter, i.e.,\,the Hubbard-Stratonovich decomposition coefficient of the interlayer Heisenberg interaction $J$ in each ordering channel. The leading SSB order (of the lowest energy $-3J/4$) appears either as an $s$-wave interlayer spin-singlet superconductivity (SC), described by $\langle \Delta_i\rangle\simeq \langle c_{i1}^\intercal\ii\sigma^2 c_{i2}\rangle\neq 0$, or as a $(\pi,\pi)$-momentum interlayer spin-singlet exciton condensation (EC), defined by $\langle \Phi_i\rangle\simeq (-)^i \langle c_{i1}^\dagger c_{i2}\rangle\neq 0$. The two SSB orders are degenerate in energy due to the $\SU(2)_1\times\SU(2)_2$ symmetry at half-filling. Both orders fully gap the Fermi surface, while breaking either the total charge $\U(1)_+$ symmetry (for the SC order) or the interlayer charge-difference $\U(1)_-$ and lattice translation symmetries (for the EC order).

\end{enumerate}

The SMG insulator can be considered a trivial insulator, achieved by restoring the broken $\U(1)_\pm$ symmetry via disordering the $\U(1)_\pm$ phase angle of the local order parameter, while still preserving the local Fermion gap. Upon (either electron or hole) doping away from the half-filling, the $\SU(2)_1\times\SU(2)_2$ symmetry is broken explicitly, such that SC and EC are no longer energetically degenerated. The  $s$-wave SC order is more favorable and becomes the unique leading SSB order, because of its perfect particle-hole Fermi surface nesting. Therefore, the high-$T_c$ SC observed in La$_3$Ni$_2$O$_7$ could be interpreted as the SC phase that arises from hole doping SMG insulator from $\nu=1/2$ to: (i) either $\nu=1/4$ for $d_{x^2-y^2}$ electrons (ii) or $\nu=1/2-\delta$ (with a small $\delta$) for $d_{z^2}$ electrons. To differentiate the two possibilities and to understand the doping effect more quantitatively, we investigate the SC phase in the lattice model \eqnref{eq: H} from both the weak coupling (BCS) and strong coupling (BEC) perspectives in the following.

\subsection{Weak Coupling Analysis}
\label{sec: weak}

\begin{figure}[htbp]
\begin{center}
\includegraphics[scale=0.65]{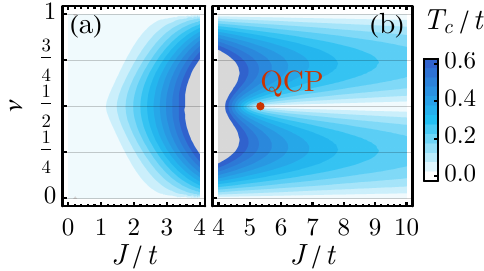}
\caption{The SC transition temperature $T_c$ depending on the interaction strength $J/t$ and the filling fraction $\nu$, estimated by (a) the BCS mean-field theory for weak coupling, and (b) the BEC mean-field theory for strong coupling. The gray regions indicate areas where mean-field theories are about to fail.}
\label{fig: meanfield}
\end{center}
\end{figure}

In the non-interacting limit ($J=0$), the Hamiltonian $H$ in \eqnref{eq: H} presents a free fermion band model, with a 4-fold (layer and spin) degenerated Fermi surface at generic filling. This free fermion state becomes unstable under the perturbative interaction $J$. In the weak coupling regime ($J/t\lesssim 1$), the mean-field energetic analysis in \tabref{tab: bilinear} suggests the leading instability occurs in the interlayer spin-singlet pairing channel, corresponding to the following Cooper pair creation operator
\eq{\Delta_i^\dagger=\tfrac{1}{\sqrt{2}}(c_{i1\uparrow}^\dagger c_{i2\downarrow}^\dagger-c_{i1\downarrow}^\dagger c_{i2\uparrow}^\dagger).}
Given the decomposition $\vect{S}_{i1}\cdot\vect{S}_{i2}=-\frac{1}{2}(\Delta_i^\dagger\Delta_i+\Delta_i\Delta_i^\dagger)+\frac{1}{4}((n_{i1}-1)(n_{i2}-1)+1)$ (where $n_{il}:=c_{il}^\dagger c_{il}$), we can formulate a Bardeen-Cooper-Shriffer (BCS) mean-field theory assuming an $s$-wave order parameter $\langle \Delta_i\rangle=\Delta$,
\eqs{H_\text{BCS}[\Delta]=\sum_{\vect{k}}&\Big(\sum_{l}c_{\vect{k}l}^\dagger (\epsilon_\vect{k}-\mu) c_{\vect{k}l}\\
&-J\big(\tfrac{1}{\sqrt{2}}\Delta c_{\vect{k}1}^\dagger \ii \sigma^2 c_{\vect{k}2}^\dagger+\text{h.c.}-|\Delta|^2\big)\Big),}
where $\epsilon_\vect{k}=2t(\cos k_x+\cos k_y)$ models the band dispersion.
The optimal $\Delta$ emerges from minimizing the mean-field free energy $F_\text{BCS}[\Delta]=-\frac{1}{\beta}\ln\Tr \e^{-\beta H_\text{BCS}[\Delta]}$. Taking the saddle point equation $\delta F_\text{BCS}[\Delta]/\delta\Delta=0$ at the $\Delta\to 0$ limit, we can compute the SC transition temperature $T_c$ (setting $k_B=1$) from
\eq{\frac{1}{J}=\int\frac{\dd^2\vect{k}}{(2\pi)^2}\frac{1}{2\xi_\vect{k}}\tanh\frac{\xi_\vect{k}}{2 T_c},}
and determine the fermion filling $\nu$ at $T_c$ by $\nu=\frac{1}{2}(1-\int\frac{\dd^2\vect{k}}{(2\pi)^2}\tanh\frac{\xi_\vect{k}}{2 T_c})$ given $\xi_\vect{k}:=\epsilon_\vect{k}-\mu$. The result is shown in \figref{fig: meanfield}(a). 

In the simplest form, the BCS theory predicts that $T_c$ should scale with the coupling strength $J$ as $T_c\propto \e^{-1/g_\text{F}J}$, where $g_\text{F}$ denotes the Fermi surface density of state (DOS). The $T_c$ vanishes either by $J=0$ or by $g_\text{F}=0$ (along $\nu=0,1$). The maximum $T_c$ occurs around $\nu=1/2$, due to the divergent DOS at the half-filling. These basic features are all manifest in \figref{fig: meanfield}(a).

\subsection{Strong Coupling Analysis}
\label{sec: strong}

We continue to consider the strong coupling mechanism, where SC emerges from doping the SMG insulator. Given that the SMG insulator already has preformed interlayer Cooper pairs in a Mott insulating state at half-filling (one Cooper pair per site), doping enables these Cooper pairs to gain mobility and condense into a superfluid (SF) state at low temperatures, manifesting as an SC state for electrons. 

In the strong coupling regime ($J/t\gg 1$), we postulate the low-energy subspace to be the Hilbert space of Cooper pairs, denoted as $\scH^\Delta=\bigotimes_i\scH_i^\Delta$, where \eq{\scH_i^\Delta=\text{span}\{\ket{0}_c, \Delta_i^\dagger\ket{0}_c, (\Delta_i^\dagger)^2\ket{0}_c\}} defines the on-site Hilbert space \footnote{Note that each site (with two layers) can host at most two Cooper pairs}. By reducing the model Hamiltonian $H$ in \eqnref{eq: H} to the low-energy subspace $\scH^\Delta$ up to the 2nd order perturbation in $t/J$, we arrive at an effective Hamiltonian $\tilde{H}$ for Cooper pairs
\eq{\label{eq: HC}
\tilde{H}=-\tilde{t}\sum_{\langle ij\rangle}(\Delta_i^\dagger \Delta_j+\text{h.c.})-\sum_{i}(2\mu N_{i-}+\tfrac{3}{4}JN_{i+}).}
where $\tilde{t}=8t^2/3J$ is the effective hopping of Cooper pairs, and $N_{i\pm}:=\Delta_i^\dagger\Delta_i\pm\Delta_i\Delta_i^\dagger$ are pair interaction ($N_{i+}$) and pair density ($N_{i-}$) operators. The Cooper pair chemical potential $\mu$ will be tuned to adjust the electron filling fraction $\nu$ effectively.

To proceed, we introduce the mean-field order parameter $\langle \Delta_i\rangle=\Delta$ to decouple the Hamiltonian $\tilde{H}$ to each site, as
\eqs{H_\text{BEC}[\Delta]=-\sum_i\Big(&4\tilde{t}(\Delta \Delta_i^\dagger+\text{h.c.}-|\Delta|^2)\\
&+2\mu N_{i-}+\tfrac{3}{4}JN_{i+}\Big).} 
The ground state energy $E_\text{BEC}[\Delta]$ of the mean-field Hamiltonian $H_\text{BEC}[\Delta]$ can be found by diagonalizing the many-body problem on each site independently (in a 3-dimensional on-site Hilbert space). The vacuum expectation value $\Delta$ can be obtained by solving the saddle point equation $\delta E_\text{BEC}[\Delta]/\delta \Delta=0$ to minimize the ground state energy $E_\text{BEC}$. The solution of $\Delta$ determines the SF density $\rho_s=|\Delta|^2$ at zero-temperature ($T=0$). Upon raising to a finite temperature ($T>0$), the broken $\U(1)_+$ charge conservation symmetry will be restored by thermal fluctuations. The transition temperature can be estimated from the Kosterlitz-Thouless transition \cite{Kosterlitz1973O} temperature $T_c=\pi \tilde{t}\rho_s=(8\pi t^2/3J)|\Delta|^2$. The result is shown in \figref{fig: meanfield}(b).

In the large $J/t$ limit, the SC transition temperature $T_c$ vanishes at $\nu=0,1/2,1$, marking different Mott insulating states of Cooper pairs, with $0,1,2$ Cooper pairs per unit cell, respectively. In terms of electrons, $\nu=0,1$ corresponds to band insulators, where $T_c$ vanishes for all $J/t$ values, owing to the absence of itinerant charge carriers. The insulating state at $\nu=1/2$ is particularly noteworthy, which corresponds to the SMG insulator, a featureless Mott insulator only attainable in bilayer lattice systems for spin-1/2 fermions, 
as a result of anomaly cancellation. 

The BEC mean-field analysis reveals a quantum critical point (QCP) at $J_c/t=16/3\approx 5.33$, where the SMG insulating phase ends, giving way to the SC phase. This QCP represents an SF-Mott transition for Cooper pairs and belongs to the (2+1)D XY universality class. Crossing this QCP (along $\nu=1/2$ with an increase in $J/t$), the electron single-particle excitation gap persists, while the phase coherence of Cooper pairs in the SC phase gets destroyed in the SMG phase, restoring the broken $\U(1)_+$  symmetry.

A key characteristic of the SMG insulator is that it is filling-enforced: since the necessary condition for the SMG to occur is anomaly cancellation, which takes place only at half-filling. As a result, if the SMG insulator is doped away from half-filling, it cannot remain featureless. The BEC mean-field analysis reveals that a superconducting (SC) order is promptly established by either electron or hole doping from $\nu=1/2$. The peak of $T_c$ is reached around $\nu=1/2\pm 1/4$ fillings, where the mobility of the Cooper pairs is at its strongest. These attributes are demonstrated in \figref{fig: meanfield}(b).

\subsection{BCS-BEC Crossover}
\label{sec: crossover}

\begin{figure}[htbp]
\begin{center}
\includegraphics[scale=0.65]{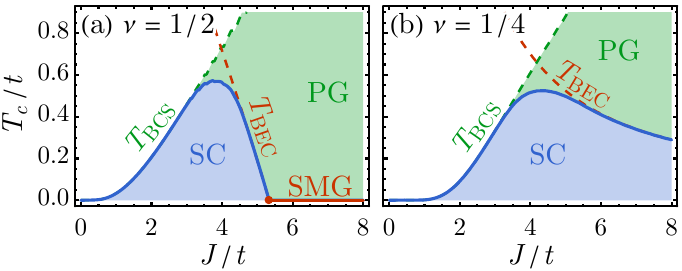}
\caption{SC transition temperature $T_c$ estimated by a smooth interpolation between BCS and BEC theories, for (a) $\nu=1/2$ and (b) $\nu=1/4$ fillings, corresponding to two horizontal cuts in \figref{fig: phase}(b). On the BEC side, a pseudo-gap (PG) phase appears above $T_c$ with incoherent preformed Cooper pairs fluctuating at low energy.}
\label{fig: crossover}
\end{center}
\end{figure}

The weak-coupling BCS and strong-coupling BEC analysis both identify the same s-wave interlayer spin-singlet SC state, implying the existence of a unified SC phase that crosses over between the BCS and BEC regimes. To qualitatively unify these two pictures, as presented in \figref{fig: meanfield}(a,b), into a single phase diagram, we employ a simple-minded smooth interpolation to estimate the SC transition temperature as:
\eq{\label{eq: interpolation}
T_c=\tfrac{T_\text{BCS}+T_\text{BEC}}{2}-T_0\ln\Big(2\cosh\big(\tfrac{T_\text{BCS}-T_\text{BEC}}{2T_0}\big)\Big).}
Here, $T_\text{BCS}$ and $T_\text{BEC}$ represent the critical temperatures predicted by the BCS and BEC theories, respectively, and $T_0$ is a temperature scale governing the smoothness of the interpolation. When $T_0=0$, \eqnref{eq: interpolation} reduces to $T_c=\min(T_\text{BCS},T_\text{BEC})$. We choose $T_0\simeq 0.067t$, which gives a reasonable interpolation as shown in \figref{fig: crossover}. The phase diagram in \figref{fig: phase}(b) is also produced with this interpolation scheme. A more rigorous calculation of $T_c$ could be performed using quantum Monte Carlo (QMC) or variational Monte Carlo (VMC) simulations, but we will leave them for future research.

As the interaction strength $J/t$ increases, the critical temperature $T_c$ exhibits a behavior that first increases (in the BCS regime) and then decreases (in the BEC regime). The BCS-BEC crossover occurs between these two regimes, where $T_c$ reaches its maximum value. The decline of $T_c$ within the BEC regime is associated with the decreasing effective hopping $\tilde{t}=8t^2/3J$ of Cooper pairs, which hinders the ability of Cooper pairs to establish phase coherence. At $\nu=1/2$, as illustrated in \figref{fig: crossover}(a), the $T_c$ drops to zero at the QCP $J_c/t=16/3$, where the SMG insulator emerges. For other fillings, such as $\nu=1/4$, the SMG phase is forbidden by the Fermi surface anomaly, so the SC phase theoretically extends to the $J/t\to \infty$ limit, as in \figref{fig: crossover}(b).

$T_{\text{BCS}}$ defines the temperature scale below which the interlayer Cooper pair forms. In the strong coupling regime, however, the Cooper pair will not establish the phase coherence and condense into the SF state, unless the temperature is lowered further below $T_{\text{BEC}}$. Thus, for large $J/t$, there exists an intermediate temperature range $T_{\text{BEC}} < T < T_{\text{BCS}}$, where the system resides in a pseudo-gap (PG) phase. The PG phase is characterized by locally fluctuating preformed Cooper pairs that lack long-range phase coherence. Itinerant electrons or holes scattering with these low-energy fluctuations can give rise to anomalous metallic behavior, such as resistivity that is linearly proportional to temperature, known as the strange metal behavior.

\begin{figure}[htbp]
\begin{center}
\includegraphics[scale=0.65]{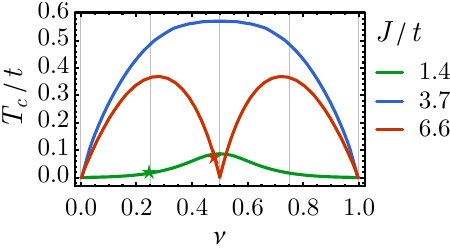}
\caption{Filling dependence of the SC transition temperature $T_c$ at different values of $J/t$, corresponding to three vertical cuts in \figref{fig: phase}(b): green - $J/t=1.4$ in the BCS regime, relevant to $d_{x^2-y^2}$ electrons; blue - $J/t=3.7$ in the crossover regime, where $T_c$ reaches global maximum; red - $J/t=6.6$ in the BEC regime, relevant to $d_{z^2}$ electrons.}
\label{fig: doping}
\end{center}
\end{figure}

At a fixed interaction strength $J/t$, the relationship between the critical temperature $T_c$ and the electron filling fraction $\nu$ is illustrated in \figref{fig: doping}. In the weak-coupling BCS regime (such as $J/t=1.4$), the maximal $T_c$ is found at half-filling due to the divergent DOS at that point. In contrast, in the strong-coupling BEC regime (such as $J/t=6.6$), the emergence of the SMG insulator at half-filling forces $T_c$ to vanish at $\nu=1/2$ and pushes the maximum $T_c$ position towards $\nu=1/4$ (or $3/4$) as $J/t\to\infty$, corresponding to 50\% hole (or electron) doping away from the SMG insulator. Clearly, doping the SMG insulator is an effective approach to induce high-temperature superconductivity, as only a small amount of doping could lead to a substantial increase in $T_c$, as shown in \figref{fig: doping} (red curve). However, the highest $T_c$ is actually reached in the BCS-BEC crossover region within this model around $J/t=3.7$, although experimentally attaining this region in real materials might be challenging.

\section{Discussions}

\subsection{Application to Pressurized La$_3$Ni$_2$O$_7$}
\label{sec: application}

We explore the high-$T_c$ SC behavior of the pressurized La$_3$Ni$_2$O$_7$ based on the mean-field understanding of the bilayer square lattice model \eqnref{eq: H}, as summarized in the phase diagram \figref{fig: phase}(b). \tabref{tab: compare} enumerates our estimates for the model parameters relevant to the Ni $d_{x^2-y^2}$ and $d_{z^2}$ electrons. 

\begin{table}[htp]
\caption{Parameter estimation and theory predicted $T_c$ for Ni $d_{x^2-y^2}$ and $d_{z^2}$ electrons. Meaning of symbols: $\nu$ - filling fraction, $t$ - hoping parameter, $J$ - interlayer Heisenberg coupling strength, $T_c$ - estimated SC transition temperature.}
\begin{center}
\begin{tabular}{l|cc}
\hline
& $d_{x^2-y^2}$ electron & $d_{z^2}$ electron\\
\hline
$\nu$ & $1/4$ & $1/2\to 0.483$ \\
\hline
$t$ & $0.48\eV$ & $\sim 0.1\eV$ \\
$J$ & \multicolumn{2}{c}{$\sim 0.66\eV$}\\
$J/t$ & $1.4$ & $6.6$ \\
\hline
$T_c/t$ (theory) & $0.014$ & $0\to 0.067$ \\
$T_c$ (theory) & $79\,\mathrm{K}$ & $0\to78\,\mathrm{K}$ \\
\hline
SC order & \multicolumn{2}{c}{$s$-wave interlayer spin-singlet}\\
Mechanism & BCS & BEC\\
\hline
\end{tabular}
\end{center}
\label{tab: compare}
\end{table}

According to these estimates,
\begin{enumerate}[(i)]
\item The itinerant $d_{x^2-y^2}$ electrons are likely located at $(J/t,\nu)=(1.4,1/4)$ in the SC phase on the BCS side, marked as the green star in \figref{fig: phase}(b), where the BCS mean-field theory predicts a SC transition temperature of the order $T_c\simeq 79\,\mathrm{K}$.

\item In contrast, $d_{z^2}$ electrons are natively positioned at $(J/t,\nu)=(6.6,1/2)$, right in the SMG insulating phase, which does not contribute to SC (i.e.,\,$T_c=0$). However, given the sensitivity of the SMG insulator to the filling fraction, an infinitesimal amount of doping can immediately drive the SMG insulator into a superconductor. 

We demonstrate that hole doping this SMG insulator by as small as $3.4\%$ to $(J/t,\nu)=(6.6,0.483)$ can place $d_{z^2}$ electrons in the SC phase on the BEC side, denoted by the red star in \figref{fig: phase}(b), with an SC transition temperature drastically increased to $T_c\simeq 78\,\mathrm{K}$, predicted by the BEC mean-field theory. DFT calculations \cite{Sun2023S2305.09586, Luo2023B2305.15564} indicate that after applying pressure, the $d_{z^2}$ band top may surpass the Fermi level, forming a smaller hole pocket, which lends support for a small amount of hole doping.
\end{enumerate}

Therefore, both $d_{x^2-y^2}$ and $d_{z^2}$ electrons can achieve high-$T_c$ SC states within the bilayer square lattice minimal model. The SC order involves $s$-wave interlayer spin-singlet pairing, driven by the interlayer antiferromagnetic Heisenberg interaction of electrons. In the phase diagram, their SC states belong to the same SC phase that emerges from doping the SMG insulator (this notion might be more appropriate for $d_{z^2}$ electrons on the BEC side). It is worth noting that the observed strange metal behavior in experiments \cite{Sun2023S2305.09586, Liu2023E2307.02950, Zhang2023H2307.14819} can be interpreted through the interaction between the two degrees of freedom: the $d_{z^2}$ electrons first form preformed incoherent Cooper pairs, which then scatter with the itinerant electrons/holes in the system, leading to the strange metal behavior. The itinerant freedom in this context may originate from either $d_{x^2-y^2}$ or $d_{z^2}$ orbitals.

Judging solely from the SC transition temperature, it is challenging to determine whether the SC in La$_3$Ni$_2$O$_7$ system originates from $d_{z^2}$ or $d_{x^2-y^2}$ electrons, as both can result in an SC transition temperature close to 80K within a reasonable parameter range. Furthermore, the strange metal behavior neither suffice to distinguish between the two cases. This uncertainty arises because it is unclear which temperature is higher between the two: the BEC temperature of the preformed pairs of $d_{z^2}$ electrons or the pairing temperature of the $d_{x^2-y^2}$ electrons. Therefore, additional experimental evidence should be considered.

On the one hand, the phenomenon of $T_c$ decreasing with increasing pressure \cite{Sun2023S2305.09586, Hou2023E2307.09865} favors the $d_{z^2}$ as the dominant degree of freedom responsible for SC. In fact, as the pressure increases, the distance between the NiO$_2$ layers decreases, and the interlayer superexchange interaction $J$ strengthens. If SC is in the BEC regime, the $T_c$ will indeed decrease with the increase in $J$. This pressure effect is harder to understand if SC is in the BCS regime. 

On the other hand, there are also compelling arguments in favor of the $d_{x^2-y^2}$ as the dominant role for SC. Firstly, some studies \cite{Cao2023F2307.06806} have shown that Hund's rule interaction has a strong renormalization effect on the $d_{z^2}$ band, leading to a significant reduction in its bandwidth and a remarkable decrease in its superconducting transition temperature. In contrast, $d_{x^2-y^2}$ has better itinerancy, making it a better candidate for SC to emerge. Secondly, for the SC resulting from doping SMG insulator, one of the most prominent features is that the optimal $T_c$ is achieved around quarter filling. This optimal filling nicely coincides with the filling of the $d_{x^2-y^2}$ orbitals, speaking for its potential of realizing high-$T_c$ SC. Finally, although the $J/t$ value of the $d_{x^2-y^2}$ orbital listed in Table \ref{tab: compare} is only 1.4, considering the no-double-occupancy constraint brought by the strong Hubbard interaction (see below), the effective $t$ would be renormalized by a Gutzwiller factor of about $\delta=0.5$, under which the value of $J/t$ doubles, which can significantly enhance the $T_c$. 

In summary, there might be simultaneous superconductivity in both $d_{z^2}$ and $d_{x^2-y^2}$ orbitals in the La$_3$Ni$_2$O$_7$ system. We propose that a key piece of evidence to distinguish between them might lie in studying the dependence of $T_c$ on doping. If the SC is dominated by $d_{x^2-y^2}$ electrons under the BCS mechanism, its $T_c$ will increase with \emph{electron} doping. If the SC is dominated by $d_{z^2}$ electrons under the BEC mechanism, its $T_c$ will increase with \emph{hole} doping. These two diametrically opposite behaviors can provide a clear distinction in determining the electronic orbital origin of SC in pressurized La$_3$Ni$_2$O$_7$ systems. Moreover, the feature of the phase transition from SC to the normal state is different between the BEC and BCS pictures. On the BEC side, the phase transition is driven by kinetic-energy gain. Consequently, the spectrum weight in the optical conductivity will shift toward the low-frequency side. Such spectrum weight shifts in the optical conductivity will not be detected in the BCS transition.

\subsection{Hubbard Interaction and Strong Correlation Effects}
\label{sec: Hubbard}

In our minimal model, we ignored the on-site Hubbard interaction among $d$-orbital electrons. Including the Hubbard interaction would amplify the electron correlation effects, manifesting as intralayer antiferromagnetic Heisenberg interaction through the superexchange mechanism. This promotes antiferromagnetic (AFM) fluctuations within each layer, potentially leading to long-range AFM order or intralayer $d$-wave SC, reminiscent of the physics in cuprates. Generally, these SSB-ordered phases could compete with the interlayer SC, leading to alternative phases \cite{Bohrdt2021E2107.08043, Bohrdt2022S2108.04118, Hirthe2023M2203.10027, Lu2023I2307.14965}.

However, analyses \cite{Lu2023I2307.14965, Oh2023T2307.15706} indicated that, for both $d_{z^2}$ and $d_{x^2-y^2}$ electrons, the intralayer AFM coupling is weaker than the interlayer coupling. Within this parameter regime, intralayer AFM correlations actually facilitate the interlayer pairing for superconductivity. Studies \cite{Chen2018T1808.06173, Zhang2022P2212.06170} on a two-leg ladder $t$-$J$ model with both intra- and inter-chain Heisenberg interactions reveal that, upon hole doping, the system develops SC quasi-long-range order with inter-chain spin-singlet pairing. The mobility of a single hole on a short-range AFM-correlated chain is hindered due to the phase string effect \cite{Sheng1996P, Weng1997Pcond-mat/9612040, Weng2011S1105.3027}, which refers to a $\dsZ_2$ Berry phase accumulated when the hole exchanges with specific background spins. This destructive interference effect reduces hole mobility and acts against superconductivity. However, because of the $\dsZ_2$ nature of the phase string effect, it is canceled if two holes across the chains are paired and move together. This results in the promotion of inter-chain pairing by the short-range AFM correlation along the chain. Such kinetic-energy-driven pairing mechanism has been studied in the bilayer square lattice very recently \cite{Zhang2023S2309.05726}, indicating an enhanced interlayer spin-singlet pairing from the phase string effect in the presence of Hubbard interaction.

\section{Summary}
\label{sec: summary}

We investigate the emergent superconductivity from doping an SMG insulator in a bilayer square lattice model, which encompasses intralayer hopping $t$ and interlayer antiferromagnetic Heisenberg interaction $J$ of spin-1/2 fermions. This minimal model is proposed to capture the essential physics of the Ni $d$-orbital electrons in the nickelate superconductor La$_3$Ni$_2$O$_7$.

We highlight the manifestation of an SMG insulator in the strong coupling $J/t\gg 1$ regime at half-filling --- a featureless Mott insulator facilitated by the cancellation of LSM anomaly in bilayer lattice systems. Our analysis reveals that doping the SMG insulator can lead to superconductivity, with an $s$-wave interlayer spin-singlet pairing. The SC phase exhibits BCS-BEC crossover by tuning the $J/t$ ratio. Our main result is the mean-field phase diagram presented in \figref{fig: phase}(b), providing a unified understanding of the SC originated from both Ni $3d_{z^2}$ and $3d_{x^2-y^2}$ orbitals.

On the strong-coupling BEC side, the SC transition temperature $T_c$ reduces with the interlayer interaction strength $J$, which provides insights into the decreasing $T_c$ with pressure in the La$_3$Ni$_2$O$_7$ superconductor. Above $T_c$, a pseudo-gap phase emerges, in which strange metal behavior is expected, consistent with experimental observation. Furthermore, our findings suggest that both Ni $3d_{z^2}$ and $3d_{x^2-y^2}$ orbitals can exhibit SC in presurized La$_3$Ni$_2$O$_7$. We propose future experimental studies of the doping dependence of the critical temperature $T_c$ or the spectral weight transfer in optical conductivity, which could likely provide experimental tests to determine which electron orbital mainly contributes to the superconducting formation.

\begin{acknowledgments}
We acknowledge insightful discussions with Meng Wang, Yi-Fan Jiang, Zi-Xiang Li, Zheyan Wan, Zheng-Yu Weng, Hao-Kai Zhang, Hui Zhai, and Yang Qi. M.L., Z.Y.Z., and Y.Z.Y. acknowledge the hospitality of Institute for Advanced Study at Tsinghua University, where part of the research was performed. J.W. thanks the hospitality of IBM Thomas J. Watson Research Center and Guanyu Zhu.
D.C.L., W.H., and Y.Z.Y. are supported by the National Science Foundation (NSF) Grant No. DMR-2238360. F.Y is supported by the National Natural Science Foundation of China under the Grant No. 12074031, No. 12234016, No. 11674025. J.W. is supported by the Center for Mathematical Sciences and Applications at Harvard University and the NSF Grant DMS-1607871. We acknowledge the OpenAI GPT4 model for providing language suggestions in writing this paper.

\end{acknowledgments}

\bibliography{ref}

\end{document}